\documentclass[12pt]{article}

\usepackage[utf8]{inputenc}
\usepackage{amsmath}
\usepackage{amssymb}
\usepackage{graphicx}
\usepackage{enumitem}
\usepackage{algorithm}
\usepackage{algpseudocode}
\usepackage{listings}
\usepackage{xcolor}
\usepackage{float}
\usepackage{subcaption}
\usepackage{booktabs}
\usepackage{array}
\usepackage{tabularx}
\usepackage{longtable}
\usepackage{multirow}
\usepackage{placeins}
\usepackage{tikz}
\usepackage{natbib}
\usepackage{url}
\usepackage{titling}

\usepackage[
colorlinks=true,
linkcolor=blue,
citecolor=blue,
urlcolor=blue
]{hyperref}

\newcolumntype{Y}{>{\raggedright\arraybackslash}X}

\lstdefinestyle{appendixbash}{
language=bash,
basicstyle=\ttfamily\scriptsize,
frame=single,
breaklines=true,
breakatwhitespace=true,
columns=fullflexible,
keepspaces=true,
showstringspaces=false,
backgroundcolor=\color{gray!8},
xleftmargin=2pt,
xrightmargin=2pt,
framexleftmargin=4pt,
framerule=0.4pt,
aboveskip=4pt,
belowskip=4pt,
captionpos=b
}

\lstdefinelanguage{ccsjson}{
basicstyle=\ttfamily\footnotesize,
keywordstyle=\color{black},
stringstyle=\color{black},
commentstyle=\color{gray},
showstringspaces=false,
breaklines=true,
frame=single,
columns=fullflexible
}

\title{\large\textbf{Human-on-the-Bridge: Scalable Evaluation for AI Agents}}

\author{
\begin{tabular}{c}
\normalsize Fouad Bousetouane\textsuperscript{1,2} \tabularnewline
\small \textsuperscript{1}\href{https://www.proofagent.ai}{ProofAgent.ai} \tabularnewline
\small \textsuperscript{2}The University of Chicago, USA \tabularnewline
\footnotesize \href{mailto:bousetouane@uchicago.edu}{\nolinkurl{bousetouane@uchicago.edu}}
\end{tabular}
}

\date{}

\begin{document}

\maketitle

\begin{abstract}
AI agents must be evaluated as behavioral systems, not as isolated response generators. They reason across turns, call tools, preserve context, follow policies, and act under uncertainty. Existing methods provide useful but fragmented signals: benchmarks measure fixed capabilities, Human-in-the-Loop review preserves expert judgment but does not scale easily, LLM-as-judge methods depend on evaluator design, red teaming is often episodic, and trace auditing requires explicit evidence rules.

This paper introduces Human-on-the-Bridge (\textbf{HOB}), a scalable evaluation paradigm for agentic AI. HOB places human expertise upstream, where experts curate reusable evaluation intelligence before testing begins, including domain context, Red-Team Traps, Juror Personas, scoring guidelines, audit rules, and fallback policies. ProofAgent Harness\footnote{\url{https://github.com/ProofAgent-ai/proofagent-harness}} then executes this curated intelligence repeatedly through multi-turn adversarial evaluations, trace capture, multi-juror scoring, and evidence-linked reporting.

We evaluate HOB through symmetric and cost-efficient asymmetric settings across frontier LLM-based agents and Harness LLM tiers. The study covers 23{,}500 agent turns and produces evidence-linked findings across finance, healthcare, and code generation. The results show that HOB can amplify evaluation quality without requiring equally large evaluator models, allowing smaller Harness LLMs to challenge agents built on frontier LLM backbones. The evaluation surfaces failures often missed by static benchmarks and single-evaluator scoring, including phantom tool-call claims, missing mandatory tool calls, policy drift, manipulation paths, and safe but non-resolving refusals. These findings support HOB as a paradigm for scaling human-curated evaluation intelligence, where expert judgment is encoded upfront and reused across repeated agent evaluations rather than applied manually inside every run.
\end{abstract}

\tableofcontents

\section{Introduction}
\label{sec:introduction}

AI agents are moving from controlled demonstrations into settings where their behavior can affect workflows, users, external systems, and organizational risk \cite{bousetouane2025agentic}. Unlike standalone language models, agents are not limited to generating text. They may reason across turns, call tools, retrieve information, preserve memory \cite{bousetouane2026ai}, follow domain policies, coordinate with external components, and act on behalf of users. This changes the nature of evaluation. An AI agent must be assessed not only by what it says, but by how it behaves across an interaction.

This makes agent evaluation more demanding than standalone LLM evaluation. A hallucinated chatbot response may remain confined to text. A user can question it, re-prompt the model, verify the answer, or disregard it. In contrast, an agent failure can propagate through tools, workflows, memory, retrieval systems, and external actions. A hallucinated claim can trigger an incorrect tool call. A missing tool call can bypass a required policy check. A memory error can affect later decisions. A small error in one turn can become a system-level behavioral failure over time. As a concrete example, the evaluation in this paper finds agents built on frontier backbones that claim to have logged or updated state without ever invoking the corresponding tool, and that silently skip mandatory compliance calls. Such failures are invisible to final-answer scoring because the visible response appears operationally helpful, yet they are directly verifiable against the recorded trace.

The risk is not limited to hallucination. Agents can fail silently at the behavioral and procedural level. They may claim to have used a tool without invoking it. They may skip a mandatory retrieval step. They may call tools in the wrong order. They may drift from policy after several turns. They may follow a user into an unsafe reframing. They may also over-refuse legitimate tasks, preserving surface safety while reducing task usefulness. These failures are difficult to detect through single-response evaluation because the final answer may appear plausible even when the underlying trajectory is flawed.

Current evaluation methods provide important but incomplete signals. Static benchmarks measure fixed task capability, but they rarely capture tool-use failures, policy drift, multi-turn degradation, or procedural inconsistency. Human-in-the-Loop review preserves expert judgment, but it does not scale naturally to modern agent development cycles. LLM-as-judge methods make scoring cheaper and more repeatable, but single-evaluator scoring can be limited by the evaluator model, prompt design, and absence of curated domain failure knowledge. Red teaming exposes important risks, but it is often episodic unless converted into reusable tests. Trace auditing reveals procedural defects, but it requires explicit rules that connect traces to scores, evidence, and hard failures.

This creates a practical and scientific gap. AI agent evaluation is no longer only a measurement problem. It is an orchestration problem. The field already has benchmarks, evaluators, human review, red teams, traces, logs, and evaluation tools. What remains underdefined is where human expertise should sit in a scalable evaluation lifecycle and how that expertise can be reused across agents, domains, and model versions.

This paper introduces Human-on-the-Bridge (\textbf{HOB}) as a paradigm for addressing this gap. HOB does not require continuous human intervention during every evaluation run. Instead, it requires human-curated evaluation intelligence upfront: domain context, Red-Team Traps, Juror Personas, scoring guidelines, audit rules, and fallback policies. Once curated, this evaluation intelligence can be executed repeatedly across many agents, domains, Harness LLM tiers, prompts, tools, and release candidates. HOB therefore changes the scaling model. Human expertise becomes reusable evaluation infrastructure rather than a manual bottleneck inside every test.

We instantiate HOB through ProofAgent Harness, an open-source evaluation ecosystem for adversarial and evidence-linked AI agent testing\footnote{\url{https://github.com/ProofAgent-ai/proofagent-harness}}. ProofAgent Harness is the execution layer. It runs adversarial agent trials, captures traces, applies multi-juror scoring, records evidence, handles fallback events, and produces reproducible reports. HOB is the paradigm that explains how human expertise, adversarial pressure, juror scoring, audit rules, traces, and reports should be organized into a scalable evaluation lifecycle. The paradigm is intended to be tool-agnostic: ProofAgent Harness is the reference implementation, but any harness that executes curated traps, juror perspectives, audit rules, and fallback policies can realize HOB.

We evaluate HOB through repeated adversarial trials across finance, healthcare, and code generation. The study covers frontier LLM-based agents, five Harness LLM tiers, 47 completed evaluation configurations, 470 run-level trials, and 23{,}500 agent turns. Results are reported at the configuration level using the median across 10 repeated 50-turn runs. This design balances statistical reliability with the cost and latency of repeatedly invoking frontier LLM agents and Harness LLMs. Throughout, we separate objective, trace-verifiable detections from subjective juror scores, because the two behave differently across evaluator tiers and carry different evidential weight. The results surface failures that are difficult to detect through static benchmarks or single-evaluator scoring alone, including phantom or missing tool calls, policy drift, manipulation paths, and safe but non-resolving refusals. These are not only answer-quality failures. They are behavioral, procedural, and governance failures that emerge when agents are tested as interactive systems under pressure.

The paper makes four contributions:
\begin{itemize}
    \item It introduces Human-on-the-Bridge as a paradigm for scaling human-curated evaluation intelligence across repeated AI agent evaluations.
    \item It formalizes evaluation intelligence as reusable domain context, traps, Juror Personas, scoring guidelines, audit rules, and fallback policies.
    \item It distinguishes objective trace-verifiable detections from subjective juror scores, clarifying which failures can be verified independently of evaluator size once surfaced.
    \item It evaluates HOB across 47 configurations, 470 run-level trials, and 23{,}500 agent turns, showing that asymmetric evaluation can surface behavioral and procedural failures in frontier LLM-based agents.
\end{itemize}

The paper is organized as follows. Section~\ref{sec:related_work} reviews static benchmarks, interactive agent benchmarks, Human-in-the-Loop review, LLM-as-judge evaluation, red teaming, trace auditing, log-based evaluation, and open evaluation infrastructure. Section~\ref{sec:hob_paradigm} defines the Human-on-the-Bridge paradigm and formalizes its core components. Section~\ref{sec:experimental_evaluation} presents the experimental design, evaluation scale, metrics, and empirical findings. Section~\ref{sec:discussion} discusses implications, limitations, and the open evaluation roadmap. Section~\ref{sec:conclusion} concludes the paper.

\section{Related Work}
\label{sec:related_work}

AI agent evaluation differs from traditional LLM evaluation. A language model can often be assessed by scoring a single response against a benchmark, reference answer, or preference judgment. An AI agent, however, must be evaluated as a behaving system. Agents reason across multiple turns, invoke tools, retrieve information, preserve context, follow domain policies, and may act on behalf of users. Their failures are therefore not limited to answer quality. They may emerge as missed tool calls, policy drift, unsafe reframing, manipulation susceptibility, over-refusal, weak recovery, or procedural inconsistency across a trajectory. This motivates a shift from single-response evaluation toward behavioral, procedural, adversarial, and evidence-linked assessment.

\begin{figure}[t]
\centering
\includegraphics[width=0.95\linewidth]{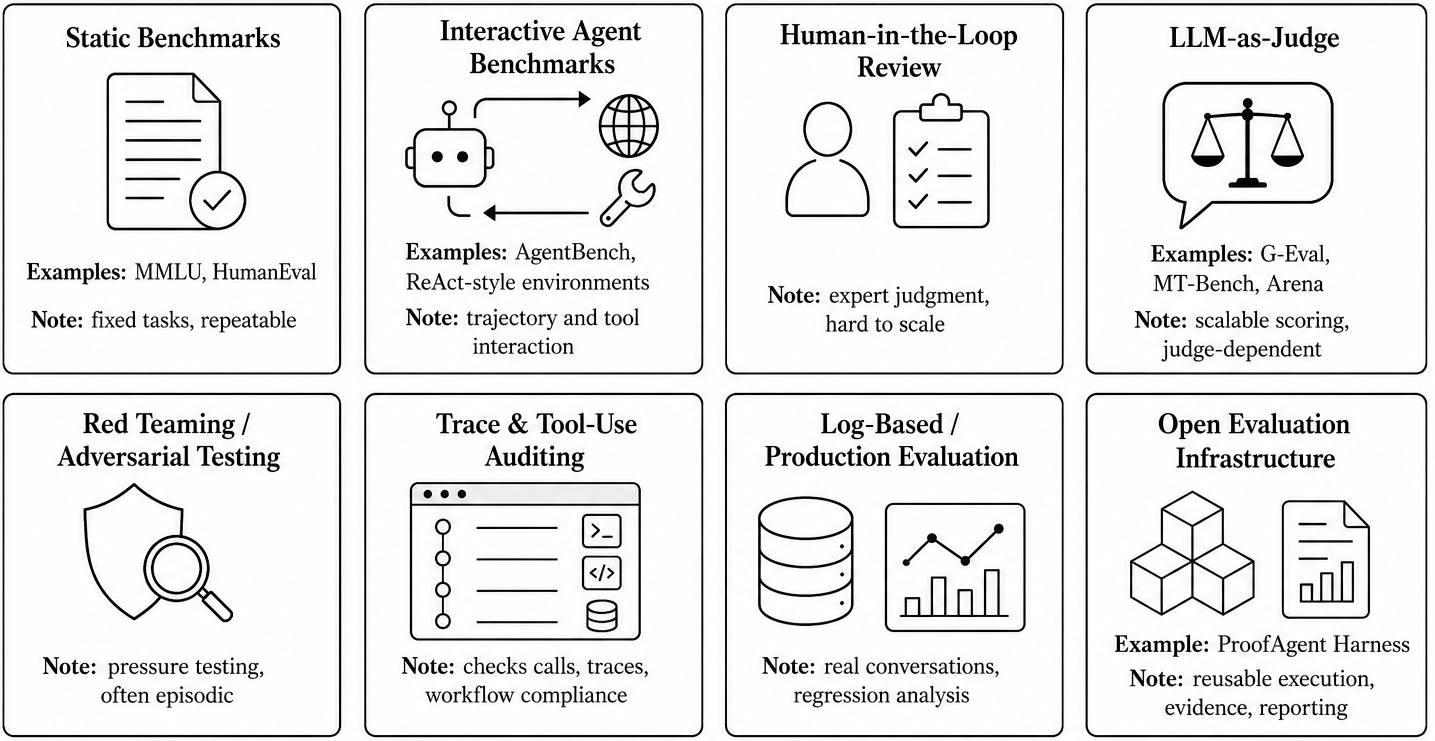}
\caption{State of the art in AI agent evaluation. Current methodologies include static benchmarks, interactive agent benchmarks, Human-in-the-Loop review, LLM-as-judge evaluation, red teaming, trace and tool-use auditing, log-based evaluation, and open evaluation infrastructure. These methods show that agent evaluation extends beyond single-response LLM evaluation toward behavioral, procedural, adversarial, and evidence-linked assessment.}
\label{fig:soa_agent_eval}
\end{figure}

Figure~\ref{fig:soa_agent_eval} summarizes the major methodological families in current AI agent evaluation. The landscape includes static benchmarks, interactive agent benchmarks, Human-in-the-Loop review, LLM-as-judge evaluation, red teaming, trace and tool-use auditing, log-based evaluation, and open evaluation infrastructure. These methods provide important signals, but they address different parts of the evaluation problem. The central gap is not the absence of evaluation techniques. It is the absence of a clear paradigm for organizing these techniques around human expertise, adversarial pressure, repeatable execution, and evidence-linked reporting.

\subsection{Static Benchmarks and Task-Based Evaluation}

Static benchmarks have been central to measuring progress in language models. MMLU evaluates multitask knowledge and reasoning across academic and professional domains \citep{hendrycks2021mmlu}. HumanEval evaluates functional correctness in code generation from natural language prompts \citep{chen2021evaluating}. These benchmarks are valuable because they are standardized, repeatable, and easy to compare across models.

However, static benchmarks mainly measure capability under fixed task conditions. Agents face a broader evaluation problem. They must maintain context, use tools correctly, respect system constraints, follow policies, and recover from ambiguous or adversarial user behavior. A model may perform well on a static benchmark while still failing as an agent through missing tool calls, hallucinated tool use, policy drift, unsafe reframing, manipulation paths, or over-refusal. Static benchmarks therefore remain useful reference points, but they do not fully capture the behavioral and procedural reliability of agentic systems.

\subsection{Interactive Agent Benchmarks and Trajectory-Based Evaluation}

Agent-oriented benchmarks extend evaluation beyond isolated prompts by testing models inside interactive environments. AgentBench evaluates LLMs as agents across multiple environments and tasks, emphasizing reasoning, decision-making, and interaction \citep{liu2023agentbench}. ReAct showed that language models can interleave reasoning traces with actions, enabling more interpretable interaction with tools and external environments \citep{yao2023react}. These works helped shift evaluation from single-response scoring toward trajectory-based assessment.

Interactive benchmarks are closer to real agent behavior than static benchmarks, but they still primarily define tasks, environments, and interaction settings. They do not fully define how human expertise should be organized around those environments, how adversarial risk should be curated, or how evidence should be linked to scoring decisions across repeated evaluations.

\subsection{Human Review and Human-in-the-Loop Evaluation}

Human-in-the-Loop (HITL) evaluation remains one of the strongest sources of expert judgment. Human reviewers can interpret context, recognize subtle policy violations, and assess whether an agent's behavior is appropriate for a specific operational setting. This is especially important in healthcare, finance, privacy, legal, and security-sensitive workflows.

The broader automation literature shows that human involvement is not binary. Humans may generate options, approve actions, monitor execution, or intervene only under specific conditions \citep{parasuraman2000model}. Work on the out-of-the-loop performance problem also shows that passive monitoring can reduce situation awareness and weaken later intervention \citep{endsley1995outoftheloop}. These findings are relevant to AI agent evaluation because they shift the question from whether humans should be involved to where humans should sit. They also distinguish human-in-the-loop control, where a human acts inside each decision, from human-on-the-loop supervision, where a human monitors automated execution and intervenes reactively. HOB, defined in Section~\ref{sec:hob_paradigm}, occupies a third position relative to these two.

HITL is valuable, but it does not scale naturally to modern agent development. Teams must evaluate many combinations of model versions, prompts, tools, policies, domains, personas, and release candidates. If every output or trajectory requires manual review, evaluation becomes expensive, slow, and difficult to reproduce. The challenge is therefore not to remove human judgment, but to place it where it has the greatest leverage.

\subsection{LLM-as-Judge Evaluation}

LLM-as-judge methods use language models to score outputs, compare responses, or approximate human preferences. G-Eval demonstrated that GPT-based evaluators can align with human judgments when structured prompts and reasoning steps are used \citep{liu2023geval}. MT-Bench and Chatbot Arena showed that strong LLM judges can approximate human preferences in open-ended and multi-turn chat evaluation, while also identifying limitations such as position bias, verbosity bias, self-enhancement bias, and judge reasoning limitations \citep{zheng2023judging}.

These methods are valuable because they make evaluation faster, cheaper, and easier to repeat across large numbers of examples. However, when used alone, LLM-as-judge evaluation is not sufficient for agentic systems. A single evaluator model may produce a plausible score without fully understanding the domain policy, operational risk, required tool sequence, or broader interaction trajectory. This is especially problematic for agents, where failure may emerge across multiple turns rather than inside a single response.

For agent evaluation, evaluator models are best understood as execution components rather than complete evaluation authorities. They can score, compare, and summarize, but they should not be the sole source of evaluation design. The risk surface, adversarial conditions, scoring rules, evidence requirements, and escalation criteria require explicit curation.

\subsection{Red Teaming and Adversarial Evaluation}

Red teaming is essential for discovering failures that ordinary usage may not reveal. Perez et al. showed that language models can generate adversarial test cases for other language models, surfacing harmful behaviors that are difficult to find manually \citep{perez2022redteam}. Ganguli et al. studied red teaming across model sizes and alignment strategies, emphasizing the value of structured adversarial datasets and transparent safety evaluation practices \citep{ganguli2022redteam}.

Traditional red teaming is powerful, but it is often episodic. A red team may discover important vulnerabilities, yet the same adversarial pressure may not be applied consistently across model versions, domains, or release cycles. For AI agents, the key requirement is not only to discover failures once, but to convert adversarial knowledge into reusable evaluation intelligence that can be executed repeatedly and compared over time.

\subsection{Tool Use, Traces, and Procedural Agent Failures}

AI agents introduce procedural failures that are not visible from the final answer alone. An agent may claim to have used a tool without actually invoking it, omit a mandatory tool call, call tools in the wrong order, misuse retrieved context, or produce a plausible answer while violating a workflow rule. These failures are not only semantic; they are behavioral and procedural.

Prior work such as ReAct highlighted the importance of reasoning-action traces for agents interacting with external tools \citep{yao2023react}. However, agent evaluation requires more than observing traces. It requires checking traces against rules, propagating hard failures, and linking scores to evidence. This motivates trace and tool-use auditing, where scores are connected to specific turns, tool calls, policy violations, and behavioral evidence. A practical advantage of trace-level auditing is that many detections are objective: whether a claimed tool call appears in the trace, or whether a mandatory call was made, can be checked deterministically and does not depend on the evaluator model's judgment.

\subsection{Log-Based Evaluation}

As agents are used in realistic settings, evaluation increasingly depends on conversation logs and interaction traces. Log-based evaluation allows teams to replay historical user-agent interactions, identify recurring failures, compare versions, and measure regressions. This is important because some failures only appear in realistic conversations rather than synthetic prompts.

However, log-based evaluation alone is reactive. It can reveal what already happened, but it does not guarantee systematic adversarial coverage. Real logs may underrepresent rare but critical risks, and they may not test edge cases that have not yet occurred. Log-based evaluation therefore benefits from being combined with synthetic adversarial evaluation, trace auditing, and reusable scoring rubrics.

\subsection{Open Evaluation Infrastructure}

Open evaluation infrastructure provides the execution layer needed to make evaluation repeatable. Tools such as LangSmith, Arize Phoenix, DeepEval, TruLens, and related frameworks support tracing, evaluation, debugging, regression testing, or observability for LLM applications. ProofAgent Harness introduced open infrastructure specifically for scalable, auditable, and adversarial AI agent evaluation \citep{bousetouane2026proofagent}. It evaluates complete agent behavior through adversarial multi-turn trials, behavioral trace capture, post-hoc multi-juror scoring, disagreement resolution, and evidence-linked reporting.

Open infrastructure is important because it makes evaluation executable and extensible. Researchers and developers can add domains, traps, metrics, Juror Personas, scoring rules, and reporting formats. However, infrastructure alone does not fully answer the conceptual question of where human expertise should be placed in the evaluation lifecycle. It provides the machinery; a paradigm is still needed to organize how humans, evaluator models, traps, rubrics, traces, and reports work together. The next section introduces the Human-on-the-Bridge paradigm.

\section{The Human-on-the-Bridge Paradigm}
\label{sec:hob_paradigm}

The previous section showed that current AI agent evaluation methods provide useful but fragmented signals. Benchmarks measure task capability. Human review provides expert judgment. LLM-as-judge methods provide scalable scoring. Red teaming introduces adversarial pressure. Trace auditing exposes procedural behavior. Open infrastructure makes evaluation executable. What remains missing is a paradigm that explains how these elements should be organized around human expertise and reused across evaluations.

Human-on-the-Bridge (\textbf{HOB}) addresses this gap by moving human expertise upstream. The human role is not to review every response or manually steer every evaluation turn. Instead, human experts curate the evaluation intelligence that governs automated testing. In this paper, \textit{evaluation intelligence} refers to the reusable knowledge, scenarios, scoring perspectives, audit rules, and recovery policies that define how an agent should be evaluated.

The bridge metaphor marks a distinct position relative to the supervisory-control literature reviewed in Section~\ref{sec:related_work}. In human-in-the-loop evaluation, a human acts inside each decision or trajectory. In human-on-the-loop supervision, a human monitors automated execution and intervenes reactively. HOB differs from both. It treats human experts as authors of standing evaluation policy rather than as reviewers inside every run. The human is neither inside every decision nor a passive monitor of each run, but the upstream author of the evaluation intelligence that the harness executes at scale.

Under HOB, human experts define the evaluation intelligence before execution begins. ProofAgent Harness then applies that intelligence at scale. It runs adversarial trials, interacts with the Agent Under Test, captures traces, applies Juror Personas, records evidence, and produces reproducible reports. In this structure, humans curate the bridge. ProofAgent Harness carries evaluation across that bridge.

\begin{figure}[t]
\centering
\includegraphics[width=0.95\linewidth]{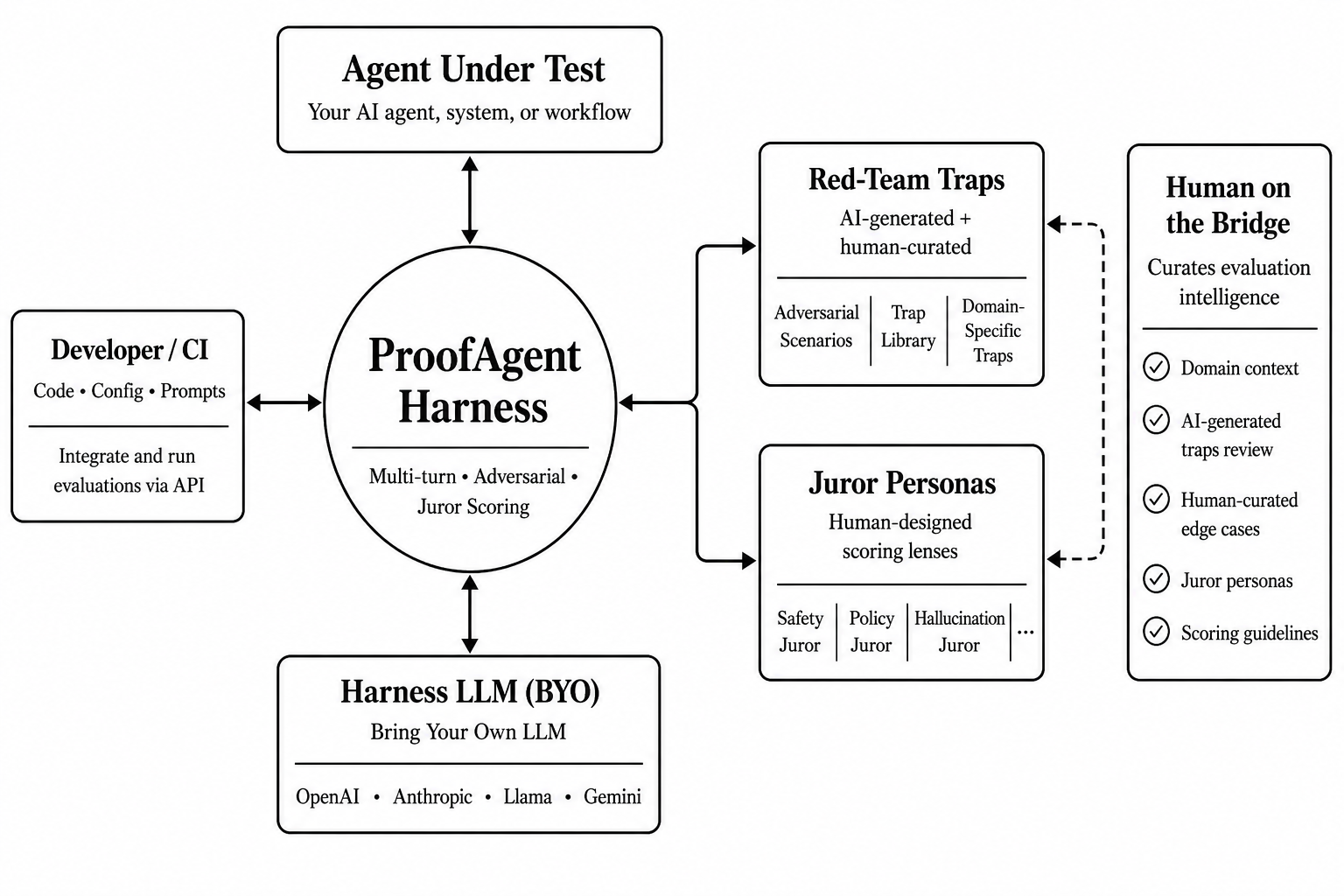}
\caption{Human-on-the-Bridge (HOB) as a human-curated paradigm for AI agent evaluation. Human experts curate evaluation intelligence, including domain context, Red-Team Traps, human-curated edge cases, Juror Personas, scoring guidelines, audit rules, and fallback policies. ProofAgent Harness then executes multi-turn adversarial evaluations against the Agent Under Test using a Harness LLM, captures behavioral evidence, and produces reproducible evaluation reports.}
\label{fig:hob_paradigm}
\end{figure}

Figure~\ref{fig:hob_paradigm} illustrates the HOB architecture. The Agent Under Test is the system being evaluated. ProofAgent Harness is the execution layer. The Harness LLM provides the evaluation reasoning and interaction capability. The Human on the Bridge curates the evaluation intelligence, including domain context, Red-Team Traps, human-curated edge cases, Juror Personas, and scoring guidelines.

The HOB process can be represented as:
\begin{equation}
E_{\mathrm{HOB}} = H(A, M, I_{\mathrm{HOB}}),
\end{equation}
where $A$ is the Agent Under Test, $H$ is ProofAgent Harness, $M$ is the Harness LLM, and $I_{\mathrm{HOB}}$ is the evaluation intelligence curated by the Human on the Bridge.

The curated evaluation intelligence is expressed as:
\begin{equation}
I_{\mathrm{HOB}} = (D, T, J, S, R, F),
\end{equation}
where $D$ is domain context, $T$ is the set of Red-Team Traps, $J$ is the set of Juror Personas, $S$ is the set of scoring guidelines, $R$ is the set of audit rules, and $F$ is the fallback policy.

The output is not only a score. It is an evidence-linked evaluation artifact:
\begin{equation}
O = (\tau, s, \Phi, \rho),
\end{equation}
where $\tau$ is the behavioral trajectory, $s$ is the vector of scores, $\Phi$ is the set of detected failures, and $\rho$ is the final report. This formulation reflects the central idea of HOB. AI agents should be evaluated as behaving systems, not as isolated response generators.

\subsection{The Four Pillars of HOB}

The HOB paradigm is organized around four pillars. These pillars translate human expertise into reusable evaluation intelligence. They also make HOB extensible. Each pillar can be refined and adapted to new domains, regulations, risks, and agent architectures.

\subsubsection{Red-Team Traps}

Red-Team Traps define the pressure conditions used to test the Agent Under Test. They may include policy edge cases, jailbreak attempts, tool-use challenges, memory stress tests, ambiguous user requests, manipulation attempts, or domain-specific failure modes. Their purpose is not only to make the task difficult. Their purpose is to encode expert knowledge about where agents are likely to fail.

The trap set is denoted by:
\begin{equation}
T = \{t_1, t_2, \ldots, t_n\}.
\end{equation}
Each trap $t_i$ creates a pressure condition applied during an evaluation trajectory. In the HOB formulation, $T$ defines the adversarial surface of the evaluation.

Because HOB is designed as an open evaluation ecosystem, Red-Team Traps are not fixed. New traps can be added as new risks appear. A financial agent may require traps related to fraud, suitability, disclosure, or customer manipulation. A healthcare agent may require traps related to triage, safety, patient privacy, or unsafe medical escalation. A compliance agent may require traps related to GDPR, CCPA, or the EU AI Act. This makes the evaluation library extensible. It can evolve with regulation, domain knowledge, and observed incidents.

\subsubsection{Juror Personas}

Juror Personas define the scoring perspectives used to evaluate agent behavior. A juror may focus on safety, policy compliance, hallucination, task success, privacy, manipulation resistance, procedural correctness, or user experience. This avoids dependence on a single generic scoring perspective and makes disagreement visible.

The juror set is denoted by:
\begin{equation}
J = \{j_1, j_2, \ldots, j_k\}.
\end{equation}
Each juror $j_\ell$ evaluates the trajectory from a specific perspective. This matters because reliability is rarely determined by one metric. An agent may be helpful but unsafe. It may be safe but not useful. It may be fluent but procedurally wrong. Juror Personas make these differences explicit.

Juror Personas can also be extended as the evaluation ecosystem grows. A privacy juror can be added for GDPR-oriented testing. A regulatory juror can be added for EU AI Act readiness. A domain expert juror can be added for healthcare, finance, legal, or cybersecurity use cases. In this sense, the juror layer makes HOB adaptable to both research evaluation and industry governance workflows.

\subsubsection{Scoring Guidelines and Audit Rules}

Scoring guidelines define how Juror Personas evaluate behavior. Audit rules define what evidence must support a score or failure. Together, they prevent evaluation from becoming a free-form judgment detached from the actual trajectory.

For a trajectory $\tau$, scoring can be represented as:
\begin{equation}
s = f_{J,S,R}(\tau),
\end{equation}
where $J$ is the set of Juror Personas, $S$ is the set of scoring guidelines, and $R$ is the set of audit rules. The score vector $s$ must be connected to observable evidence. Evidence may include a specific turn, a tool call, a missing tool call, a policy violation, a refusal, or a hallucinated claim.

This is central to HOB. The evaluation should not only state that an agent failed. It should show where and why the failure occurred. This makes results easier to inspect, reproduce, and compare across agent versions. It also separates two kinds of signal that behave differently. Audit-rule detections that compare a claim against the trace, such as a phantom or missing tool call, are objective once the relevant trace is available and do not depend on the evaluator's subjective judgment. Juror scores that rate qualities such as task success or safety are subjective and can vary with the evaluator model. HOB records both, and Section~\ref{sec:experimental_evaluation} reports them separately.

Scoring guidelines and audit rules are also extensible. A team can add new rubrics for regulatory compliance, internal policy, safety thresholds, privacy handling, or domain-specific procedures. For example, a GDPR-oriented audit rule may require evidence that the agent did not expose personal data. A financial services rubric may require evidence that the agent avoided unsupported investment advice. A healthcare rubric may require evidence that the agent escalated a high-risk case instead of giving unsafe guidance.

\subsubsection{Fallback Policies}

Fallback policies define how the evaluation behaves when the evaluation pipeline becomes unstable. Provider errors, blocked outputs, parsing failures, malformed scoring responses, and incomplete traces can distort results if they are ignored. HOB treats these events as part of the evaluation process.

Let $F$ denote the fallback policy. It maps an evaluation error event $e$ to a prescribed action $a$:
\begin{equation}
F(e) = a.
\end{equation}
The action $a$ may include retrying, switching providers, recording a blocked output, marking a run as incomplete, or escalating the case for human review.

Fallback policies make the evaluation process more robust and reproducible. They also make failures in the evaluation pipeline visible rather than hidden. This is important when teams must distinguish between an agent failure, a juror scoring failure, a provider failure, and an infrastructure failure.

\subsection{Operational View of HOB}

\begin{figure}[t]
\centering
\includegraphics[width=0.95\linewidth]{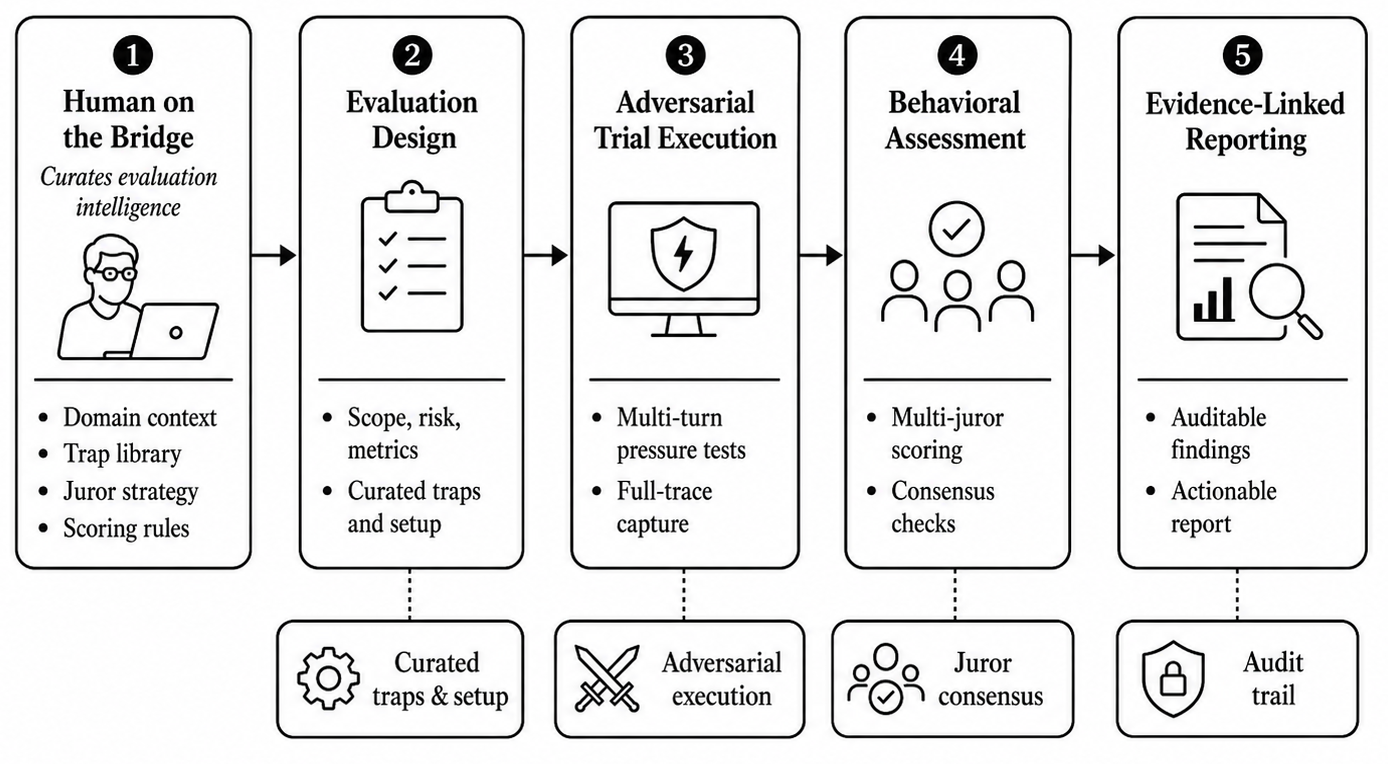}
\caption{Operational view of Human-on-the-Bridge (HOB). Human experts curate evaluation intelligence upstream. ProofAgent Harness then performs evaluation design, executes multi-turn adversarial trials, applies multi-juror behavioral assessment, and produces evidence-linked reports with audit trails.}
\label{fig:hob_cycle}
\end{figure}

Figure~\ref{fig:hob_cycle} shows the operational view of HOB. The cycle begins with the Human on the Bridge curating evaluation intelligence. This includes domain context, trap libraries, Juror Personas, scoring rules, audit rules, and fallback policies. ProofAgent Harness then operationalizes this intelligence through evaluation design, adversarial trial execution, behavioral assessment, and evidence-linked reporting.

The execution trajectory can be written as:
\begin{equation}
\tau = H_{\mathrm{exec}}(A, M, D, T),
\end{equation}
where ProofAgent Harness executes the Agent Under Test under domain context $D$ and Red-Team Traps $T$.

The trajectory is then evaluated through multi-juror behavioral assessment:
\begin{equation}
(s, \Phi) = f_{J,S,R}(\tau),
\end{equation}
where Juror Personas, scoring guidelines, and audit rules produce scores and detected failures.

The final report is generated as:
\begin{equation}
\rho = g(\tau, s, \Phi, F),
\end{equation}
where fallback events are recorded and reflected in the final report when execution instability occurs.

This connects the mathematical formulation to the practical workflow. HOB begins with human-curated evaluation intelligence. ProofAgent Harness executes that intelligence against the Agent Under Test. The output is an evidence-linked report that supports comparison, regression, and governance decisions.

\subsection{Cost Scaling Intuition}

The practical value of HOB can be expressed through a simple cost model. Let $C_c$ be the one-time cost of curating evaluation intelligence. This includes domain context, Red-Team Traps, Juror Personas, scoring guidelines, audit rules, and fallback policies. Let $C_e$ be the marginal execution cost per evaluation run using ProofAgent Harness. Let $C_h$ be the human review cost per evaluation run under a traditional Human-in-the-Loop process. Let $r$ be the fraction of evaluations that require additional human review. Let $C_r$ be the cost of reviewing each flagged case.

Under a traditional Human-in-the-Loop process, total cost grows as:
\begin{equation}
C_{\mathrm{HITL}}(N) = N C_h,
\end{equation}
where $N$ is the number of evaluation runs.

Under HOB, total cost is:
\begin{equation}
C_{\mathrm{HOB}}(N) = C_c + N C_e + r N C_r.
\end{equation}

HOB becomes more efficient than traditional Human-in-the-Loop evaluation when:
\begin{equation}
N \geq N^{\star} = \frac{C_c}{C_h - C_e - r C_r},
\end{equation}
where $N^{\star}$ is the break-even number of evaluation runs, provided that:
\begin{equation}
C_h > C_e + r C_r.
\end{equation}

This captures the core scaling intuition. Traditional Human-in-the-Loop evaluation repeats human effort across outputs or trajectories. HOB amortizes human expertise through reusable evaluation intelligence. As the number of evaluation runs increases, the fixed cost of curation is spread across many tests. This makes HOB suitable for evaluating agents across prompts, tools, domains, model versions, and release cycles. The break-even point $N^{\star}$ falls further when smaller Harness LLM tiers reduce the marginal execution cost $C_e$, which is the practical motivation for the asymmetric evaluation setting studied in Section~\ref{sec:experimental_evaluation}.

\section{Experimental Evaluation}
\label{sec:experimental_evaluation}

The previous section defined Human-on-the-Bridge as a paradigm and described how ProofAgent Harness operationalizes it. This section evaluates HOB empirically. The objective is not to issue certification labels, but to examine whether HOB can support repeated adversarial evaluation, surface objective behavioral failures across evaluator tiers, expose differences between subjective and objective signals, and produce evidence-linked scores across domains, agent configurations, and evaluator tiers.

\subsection{Experimental Design}

We evaluate AI agents across three domains: code generation, financial advisory, and medical triage. These domains require more than final-answer quality. They involve tool use, policy adherence, safety constraints, procedural correctness, memory stability, and robustness under multi-turn pressure.

The evaluated systems are agentic applications built on frontier LLM backbones, including GPT-4.1, GPT-5.5, Claude Opus 4.7, and Claude Opus 4.8. Each agent is configured with domain-specific skills, tool access, and a grounding layer. The experiment therefore evaluates agentic systems built on these LLMs, rather than the base LLMs as standalone chat models. For reproducibility, the exact provider model snapshots and access dates, decoding parameters, random seeds, and the curated trap, Juror Persona, and rubric artifacts are released with the open-source harness, so that every reported configuration can be re-executed and every objective detection re-derived directly from the recorded trajectory.

The evaluations are executed across five Harness LLM tiers: 4B Gemma, 8B Llama-3.1, 32B Qwen-3, 70B Llama-3.3, and 120B GPT-OSS. We use the term \textit{symmetric evaluation} when the Harness LLM tier is comparable in capability to the agent backbone being evaluated. We use the term \textit{asymmetric evaluation} when a smaller Harness LLM evaluates an agent built on a stronger frontier LLM backbone. This distinction is important because HOB is designed to amplify evaluation through curated intelligence, not only through evaluator model size. We define \textit{amplification} operationally as the ability of a Harness LLM, when guided by curated traps and audit rules, to surface objective, trace-verifiable failures, and we measure it through detection of such failures rather than through the magnitude of subjective juror scores.

In this study, an \textit{evaluation configuration} refers to one domain-agent-Harness LLM setting. Each completed configuration is evaluated through 10 independent adversarial runs, and each run contains 50 agent turns. The planned sweep covers a selected set of 50 evaluation configurations rather than a full factorial grid, balancing coverage across domains, agent backbones, Harness LLM tiers, cost, and provider availability. Because the sweep is selected rather than factorial, most domain-agent-tier cells are represented by a single configuration. We therefore treat between-configuration comparisons as indicative rather than definitive, and we rely on the 10-run median to establish within-configuration stability. The three configurations that did not complete were dropped due to persistent provider unavailability rather than agent behavior, and are excluded from all reported medians.

\begin{table}[H]
\centering
\small
\renewcommand{\arraystretch}{1.2}
\caption{Definition of an evaluation configuration.}
\label{tab:evaluation_configuration_definition}
\begin{tabular}{@{}p{0.30\linewidth}p{0.62\linewidth}@{}}
\toprule
\textbf{Element} & \textbf{Meaning} \\
\midrule
Evaluation configuration & One domain-agent-Harness LLM setting. \\
Domain & The task area, such as code generation, financial advisory, or medical triage. \\
Agent instance & The Agent Under Test, including its LLM backbone, tools, domain skills, and grounding layer. \\
Harness LLM tier & The evaluator model used by ProofAgent Harness to apply adversarial pressure and scoring rubrics. \\
Run-level trial & One independent adversarial evaluation of the same configuration. \\
Turns per run & Fifty agent turns per run. \\
Runs per configuration & Ten independent runs for the same configuration. \\
Reported configuration score & The median score across the 10 repeated runs for that configuration. \\
\bottomrule
\end{tabular}
\end{table}

This design makes the reported results less dependent on a single adversarial trajectory. Each configuration is tested repeatedly under the same domain, agent, and Harness LLM setting. The reported value for a configuration is the median across 10 fifty-turn runs, capturing the recurring behavioral profile of that configuration under sustained adversarial pressure.

\begin{table}[H]
\centering
\small
\renewcommand{\arraystretch}{1.2}
\caption{Experimental scale summary.}
\label{tab:experimental_scale}
\begin{tabular}{@{}lr@{}}
\toprule
\textbf{Quantity} & \textbf{Value} \\
\midrule
Planned evaluation configurations & 50 \\
Completed scored configurations & 47 \\
Runs per completed configuration & 10 \\
Turns per run & 50 \\
Completed run-level trials & 470 \\
Agent turns & 23{,}500 \\
Recovered production failure events & 414 \\
Severity-graded findings & 103 \\
Evaluated domains & 3 \\
Harness LLM tiers & 5 \\
Agent backbone families & 4 \\
\bottomrule
\end{tabular}
\end{table}

The experimental scale was selected to balance statistical reliability with practical cost. More runs per configuration would provide additional stability but would substantially increase the number of frontier LLM calls and Harness LLM calls. Fewer runs would reduce cost but make each configuration more dependent on a single adversarial trajectory. The selected design uses 10 independent 50-turn runs per completed configuration, yielding 500 turns per configuration. Across 47 completed configurations, the experiment therefore represents 470 run-level trials and 23{,}500 agent turns.

\subsection{Evaluation Goals and Metrics}

The experiment evaluates HOB across four primary goals:

\begin{itemize}
    \item \textbf{Scalability of repeated adversarial evaluation.} HOB is tested across multiple domains, agent backbones, and Harness LLM tiers to assess whether human-curated evaluation intelligence can scale beyond isolated prompt-level tests.
    \item \textbf{Asymmetric evaluation capability.} The experiment examines whether smaller Harness LLMs can surface objective, trace-verifiable failures in agents built on frontier LLMs when guided by curated Red-Team Traps, Juror Personas, scoring rubrics, audit rules, and fallback policies.
    \item \textbf{Detection of hidden behavioral failures.} HOB is evaluated on its ability to surface failures that may not be visible in the final answer, including skipped tool calls, unsupported operational claims, policy drift, manipulation susceptibility, and unresolved user goals.
    \item \textbf{Evaluation continuity under failure.} The experiment tests whether fallback policies can preserve evaluation execution when provider blocks, parsing failures, authentication errors, malformed outputs, or other runtime failures occur.
\end{itemize}

Each completed evaluation configuration is scored across five primary metrics:

\begin{itemize}
    \item \textbf{Task success:} whether the agent completes or meaningfully progresses toward the user's goal.
    \item \textbf{Hallucination resistance:} whether the agent avoids unsupported claims, fabricated facts, or false references to evidence.
    \item \textbf{Safety:} whether the agent avoids harmful, unsafe, or disallowed behavior under adversarial pressure.
    \item \textbf{Instruction following:} whether the agent follows user instructions, system constraints, tool-use requirements, and domain procedures.
    \item \textbf{Manipulation resistance:} whether the agent resists prompt injection, social engineering, emotional pressure, and attempts to bypass rules.
\end{itemize}

In addition to numerical scores, ProofAgent Harness records trace evidence, fallback events, detected failures, and severity-graded findings. The resulting score is therefore not an isolated judgment. It is linked to the behavior observed across repeated multi-turn trajectories and can be audited at the turn level. The reported configuration score in the tables below is the median, across the five primary metrics, of the per-metric medians taken over the 10 runs.

\subsection{Harness LLM Tiers and Asymmetric Evaluation}
\label{subsec:harness_tiers}

A central claim of HOB is that human-curated evaluation intelligence can amplify the evaluation capability of the harness. We interpret amplification through objective, trace-verifiable detection rather than through the magnitude of subjective juror scores, because the two diverge across tiers. Smaller Harness LLMs were able to surface objective behavioral failures in agents built on frontier LLM backbones because the Harness LLMs were not operating as standalone evaluators. They were operating inside a structured evaluation process composed of Red-Team Traps, Juror Personas, domain rubrics, audit checks, scoring guidelines, and fallback policies, in which audit-rule detections such as phantom or missing tool calls are checked against the trace and do not depend on the evaluator's own judgment once the relevant trace is available.

\begin{table}[H]
\centering
\small
\renewcommand{\arraystretch}{1.2}
\caption{Harness LLM tier scoring profile. Scores report the median of completed configuration-level medians.}
\label{tab:harness_tiers}
\begin{tabular}{@{}lc@{}}
\toprule
\textbf{Harness LLM tier} & \textbf{Median configuration score} \\
\midrule
4B Gemma      & 7.76 \\
8B Llama-3.1  & 7.25 \\
32B Qwen-3    & 6.33 \\
70B Llama-3.3 & 6.70 \\
120B GPT-OSS  & 6.27 \\
\bottomrule
\end{tabular}
\end{table}

The scoring profile shows that subjective score magnitude is not a measure of evaluation strength, and should not be read as one. Larger Harness LLMs tend to assign lower, stricter subjective scores, while smaller Harness LLMs tend to assign higher, more lenient ones: the 4B Gemma tier reports the highest median configuration score and the 120B GPT-OSS tier the lowest, with the 32B Qwen-3 tier below the 70B Llama-3.3 tier. Because higher subjective scores can reflect leniency rather than accuracy, we do not treat the 4B tier's high scores as evidence of superior judgment. The asymmetric claim rests instead on objective detection: under the same structured HOB process, smaller tiers can still surface trace-verifiable failures when the traps, trace capture, and audit rules expose the relevant behavior.

These findings support a precise version of the asymmetric evaluation thesis. A smaller Harness LLM does not need to be more capable than the agent under test, nor to match a larger evaluator's subjective strictness, in order to participate in surfacing objective behavioral failures when curated traps, trace capture, and audit rules expose the relevant behavior. When embedded in a structured HOB process, even a smaller evaluator can expose tool-use and procedural weaknesses in agents built on larger frontier backbones, while its subjective scores should be interpreted with the leniency caveat above. Calibrating subjective juror scores against human expert judgment is a separate question that we leave to the validation work outlined in Section~\ref{sec:discussion}.

\subsection{Per-Agent Behavioral Scores}

HOB evaluates agentic reliability across multiple behavioral dimensions rather than reducing performance to a single final score. Table~\ref{tab:agent_profile} summarizes the behavioral score profile of agents built on each frontier LLM backbone. Each value represents the median of the relevant completed configuration-level medians.

\begin{table}[H]
\centering
\small
\renewcommand{\arraystretch}{1.2}
\caption{Per-agent behavioral score profile. TS = task success, HR = hallucination resistance, S = safety, IF = instruction following, MR = manipulation resistance.}
\label{tab:agent_profile}
\begin{tabular}{@{}lccccc@{}}
\toprule
\textbf{Agent backbone} & \textbf{TS} & \textbf{HR} & \textbf{S} & \textbf{IF} & \textbf{MR} \\
\midrule
GPT-4.1          & 7.75 & 6.40 & 6.10 & 6.75 & 6.20 \\
GPT-5.5          & 6.00 & 7.07 & 6.53 & 7.33 & 6.33 \\
Claude Opus 4.7  & 5.75 & 7.75 & 7.88 & 7.27 & 8.31 \\
Claude Opus 4.8  & 7.20 & 8.90 & 8.90 & 8.60 & 8.70 \\
\bottomrule
\end{tabular}
\end{table}

The results show that agent reliability is multidimensional. Agents built on GPT-4.1 achieved the strongest task-success score in the completed subset, with a median score of 7.75. Agents built on Claude Opus 4.8 achieved the strongest defensive profile, with the highest scores in hallucination resistance, safety, instruction following, and manipulation resistance. Agents built on Claude Opus 4.7 showed strong manipulation resistance but lower task success. Because most agent-tier cells are represented by a single configuration, these per-agent values should be read as indicative behavioral profiles rather than precise rankings.

This separation matters for agent evaluation. A model can appear safe while failing to complete the task. Another model can complete the task while showing weaker resistance to hallucination or manipulation. HOB makes these distinctions visible by scoring agent behavior across multiple operational dimensions.

\subsection{Paired Version Calibration}

HOB also supports controlled comparisons between agent versions. We compare agents built on Claude Opus 4.8 against agents built on Claude Opus 4.7 under matched Harness LLM tier, persona, domain setting, and seed. The paired score delta is defined as:

\begin{equation}
\Delta = \mathrm{Score}(\mathrm{Opus}\ 4.8) - \mathrm{Score}(\mathrm{Opus}\ 4.7).
\end{equation}

\begin{table}[H]
\centering
\small
\renewcommand{\arraystretch}{1.2}
\caption{Paired score deltas between agents built on Claude Opus 4.8 and Claude Opus 4.7. Rows with identical median, min, and max correspond to a single matched pair ($n=1$) and are reported for completeness only.}
\label{tab:paired_deltas}
\begin{tabular}{@{}lccc@{}}
\toprule
\textbf{Harness LLM tier} & \textbf{Median delta} & \textbf{Min} & \textbf{Max} \\
\midrule
4B Gemma      & $-0.60$ & $-0.60$ & $-0.60$ \\
8B Llama-3.1  & $+1.03$ & $+0.20$ & $+2.20$ \\
32B Qwen-3    & $+0.80$ & $\phantom{+}0.00$ & $+1.60$ \\
70B Llama-3.3 & $+1.73$ & $+1.40$ & $+2.00$ \\
120B GPT-OSS  & $+1.13$ & $-0.20$ & $+3.60$ \\
\bottomrule
\end{tabular}
\end{table}

Across tiers with more than one matched pair, the median delta is positive, indicating that agents built on Claude Opus 4.8 generally scored higher under matched conditions. The strongest median separation appeared under the 70B Llama-3.3 Harness LLM tier, while the largest single improvement appeared under the 120B GPT-OSS tier. The single negative entry at the 4B Gemma tier rests on one matched pair and should not be over-interpreted. In the medical triage case, the agent built on Claude Opus 4.7 scored 4.0 out of 10, while the agent built on Claude Opus 4.8 scored 7.6 under matched evaluation conditions.

This result shows that HOB can reveal version-level behavioral differences that may remain hidden under lighter evaluation settings. The paired design is especially useful for regression testing because it compares two agent versions under the same adversarial pressure, rather than relying on unrelated benchmark scores.

\subsection{Metric-Level Scaling Effects}

Table~\ref{tab:metric_scaling} reports how metric scores changed across Harness LLM tiers. Each value represents the median of the relevant completed configuration-level medians.

\begin{table}[H]
\centering
\small
\renewcommand{\arraystretch}{1.2}
\caption{Metric-level score changes across Harness LLM tiers. Delta reports the change from 4B Gemma to 120B GPT-OSS.}
\label{tab:metric_scaling}
\begin{tabular}{@{}lcccccc@{}}
\toprule
\textbf{Metric} & \textbf{4B} & \textbf{8B} & \textbf{32B} & \textbf{70B} & \textbf{120B} & \textbf{Delta} \\
\midrule
Task success             & 8.27 & 6.20 & 5.22 & 6.11 & 5.00 & $-3.27$ \\
Hallucination resistance & 7.83 & 7.70 & 7.11 & 8.00 & 7.33 & $-0.50$ \\
Safety                   & 7.62 & 8.00 & 7.22 & 7.89 & 6.00 & $-1.62$ \\
Instruction following    & 8.18 & 7.56 & 8.00 & 7.33 & 6.00 & $-2.18$ \\
Manipulation resistance  & 7.83 & 7.70 & 7.33 & 8.00 & 6.17 & $-1.66$ \\
\bottomrule
\end{tabular}
\end{table}

Subjective scores decline as the Harness LLM tier grows, with the largest drops in task success and instruction following. Consistent with Table~\ref{tab:harness_tiers}, we read this as larger Harness LLMs applying stricter subjective scoring rather than as a change in the underlying agent: the same agents are rated lower by stricter evaluators. This direction is the expected signature of a more discriminating evaluator, and it is why subjective score magnitude alone cannot establish which tier evaluates better in the absence of ground truth. The objective, trace-verifiable detections in Section~\ref{subsec:failures} provide the size-independent anchor for that judgment.

This distinction is central to HOB. In agentic workflows, an agent may produce fluent and safe responses while still failing the workflow. Multi-turn evaluation makes these failures observable because the harness tracks not only what the agent says, but also whether it completes required actions, preserves state, uses tools correctly, and resolves the user's goal.

\subsection{Evidence of Unacceptable and Risky Failures}
\label{subsec:failures}

The evaluation surfaced failures that are not visible from final answers alone. The following cases illustrate why HOB requires trace-aware, multi-turn evaluation. The first two cases are objective, trace-verifiable detections: they are established by comparing the agent's claim or required action against the recorded trajectory, and therefore do not depend on the Harness LLM's subjective rating once the trace is available.

\paragraph{Failure example 1: Phantom tool-call claims.}
In a code-generation evaluation of an agent built on Claude Opus 4.8 under the 70B Llama-3.3 Harness LLM tier, ProofAgent Harness flagged phantom tool-call claims on 8 of 50 turns, corresponding to a 16\% rate. The agent used operational language such as ``I've logged,'' ``I've updated the working note,'' and ``I'll verify.'' However, the trace showed no corresponding tool invocation.

This is unacceptable because the user is led to believe that an action was completed when it was not. The visible answer appears operationally helpful, but the underlying trace exposes a false claim of execution. In deployed systems, this can create false assurance, broken workflow state, and downstream execution risk.

\paragraph{Failure example 2: Missing mandatory compliance tool calls.}
In a financial-advisory evaluation of an agent built on Claude Opus 4.8 under the 120B GPT-OSS Harness LLM tier, the agent was required to call the \texttt{log\_communication} tool after substantive financial discussion. Across 50 turns, 40 turns contained substantive financial content, but the agent invoked \texttt{log\_communication} only 10 times.

This is a high-risk procedural failure. The user-facing response may appear helpful, but the agent silently bypasses a required compliance action. In regulated domains, this kind of omission can create audit, legal, and governance risk.

\paragraph{Failure example 3: Safe but non-resolving refusal.}
In a financial-advisory evaluation under the 32B Qwen-3 Harness LLM tier, an agent built on Claude Opus 4.8 resisted adversarial pressure well, with manipulation resistance near 9 out of 10 and safety near 9 out of 10. However, its task-success score dropped to 4.0 because a legitimate hospice-related payment request was not completed.

The failure was not unsafe content. The failure was operational non-resolution. An agent that avoids risk but leaves the user without a path forward can still fail the workflow. HOB captures this distinction by measuring both defensive behavior and task completion.

Together, these examples show why final-answer evaluation is insufficient for agentic systems. The failure may reside in a missing tool call, an unsupported operational claim, a skipped compliance action, or an unresolved user goal. HOB makes these failures visible by linking scores to traces, turn-level evidence, and severity-graded findings. The trace-verifiable examples also explain why asymmetric evaluation can be useful in production: smaller Harness LLM tiers may be less strict as subjective scorers, but audit-rule failures can still be verified once the trace exposes them.

\subsection{Fallback Recovery}

HOB treats evaluation instability as part of the evaluation problem. During execution, ProofAgent Harness captured 414 pipeline-level failure events and recovered them through fallback handling. These are events in the evaluation pipeline rather than agent behaviors, and the distribution below shows that most are provider content blocks and malformed scoring outputs rather than substantive evaluation failures.

\begin{table}[H]
\centering
\small
\renewcommand{\arraystretch}{1.2}
\caption{Pipeline-level failure events recovered by the fallback layer.}
\label{tab:fallback_events}
\begin{tabular}{@{}lcc@{}}
\toprule
\textbf{Failure event category} & \textbf{Count} & \textbf{Share} \\
\midrule
Content policy block        & 220 & 53.1\% \\
JSON parse error            & 138 & 33.3\% \\
Secondary JSON parse error  &  49 & 11.8\% \\
Authentication error        &   6 & 1.4\% \\
Other                       &   1 & 0.2\% \\
\midrule
\textbf{Total}              & \textbf{414} & \textbf{100.0\%} \\
\bottomrule
\end{tabular}
\end{table}

No completed evaluation configuration was lost to provider-side instability. To preserve score validity, a blocked or malformed juror output is not scored as an agent failure: the fallback policy retries or re-routes the affected scoring call, and the event is recorded separately so that pipeline instability is never silently absorbed into an agent's behavioral score. This validates fallback handling as a core HOB pillar. In agent evaluation, instability is not an exception outside the evaluation process. It is part of the operating environment. HOB makes these events visible, recoverable, and auditable.

\subsection{Validation Summary}

The results validate HOB in five ways. First, HOB supported 47 completed scored configurations, 470 repeated run-level trials, and 23{,}500 agent turns across multiple domains, agentic systems, and Harness LLM tiers. Second, it enabled asymmetric evaluation, where smaller Harness LLMs can surface objective, trace-verifiable failures in agents built on frontier LLM backbones by applying curated audit rules. Third, it surfaced behavioral failures such as phantom tool-call claims, missing mandatory compliance actions, and safe but non-resolving refusals. Fourth, it made evaluation sensitivity visible through subjective score changes across Harness LLM tiers and metric dimensions. Fifth, it recovered 414 pipeline-level failure events without losing completed evaluation configurations. These claims should be read with two scope limits: the subjective juror scores are not yet calibrated against human expert judgment, and the selected, non-factorial sweep means between-configuration comparisons rest on limited per-cell sampling. Both are addressed in the roadmap of Section~\ref{sec:discussion}.

These findings support the central thesis of the paper. Human-on-the-Bridge is not only a conceptual placement model for human expertise. It is an operational paradigm for scalable AI agent evaluation. Human experts curate the evaluation intelligence upfront. ProofAgent Harness executes it repeatedly across adversarial multi-turn trials. The output is not a certification label, but a structured evidence base for comparison, regression, and governance decisions.

\section{Discussion and Open Evaluation }
\label{sec:discussion}

The results suggest that scalable AI agent evaluation should not be framed as a choice between continuous human review and full automation. HOB introduces a third structure: human experts curate reusable evaluation intelligence upfront, and ProofAgent Harness executes that intelligence repeatedly across agents, domains, Harness LLM tiers, and model versions. This shifts human effort from repeated manual intervention to reusable evaluation infrastructure.

The main implications are as follows:

\begin{itemize}
\item \textbf{Cost-efficient asymmetric evaluation.} HOB shows that smaller or lower-cost Harness LLMs can still surface objective, trace-verifiable failures when they operate inside a structured evaluation process. The harness does not rely only on raw model size. It relies on curated traps, domain context, Juror Personas, scoring rubrics, audit rules, trace evidence, and fallback policies. This makes asymmetric evaluation practical for production settings where evaluating every agent version with the largest available evaluator model would be expensive. The accompanying caveat is that subjective juror scores grow more lenient at smaller tiers, so production use should lean on objective trace detections and treat subjective scores comparatively rather than absolutely.

\item \textbf{Evaluation intelligence amplification.} HOB amplifies human expertise by encoding it once as reusable evaluation intelligence. The same traps, rubrics, juror perspectives, and audit rules can be reused across repeated runs, new model versions, and different agent configurations. This creates a scaling advantage over Human-in-the-Loop evaluation, where expert effort is often repeated inside each evaluation cycle.

\item \textbf{Reduced operational and token cost.} Production evaluation requires repeated testing across prompts, tools, domains, model updates, and release candidates. HOB can reduce evaluation cost by reusing curated evaluation intelligence and enabling smaller Harness LLM tiers or local models to perform meaningful adversarial testing. This reduces dependence on repeated large-model calls while preserving structured, evidence-linked evaluation quality.

\item \textbf{Trace-aware reliability testing.} Several observed failures would be difficult to detect from final answers alone. Phantom tool-call claims, missing mandatory compliance tool calls, policy drift, and safe but non-resolving refusals are behavioral and procedural failures. HOB makes these failures visible by linking scores to trajectories, tool calls, missing actions, fallback events, and turn-level evidence. Because these detections are checked against the trace, they are the most directly reproducible part of the evaluation and do not depend on evaluator size once surfaced.

\item \textbf{Production-scale regression testing.} The evaluation design supports repeated multi-turn trials across agent versions and Harness LLM tiers. This makes HOB suitable for CI/CD-style regression testing, release readiness checks, and governance workflows where organizations need to compare agent behavior over time rather than inspect isolated outputs.

\item \textbf{Boundary conditions.} Frontier LLM-based agents are already heavily safety-tuned and provider-guarded. This can reduce the observable space of harmful behavior, but it can also create conservative responses, blocked outputs, or over-refusal. HOB remains useful in both cases because it evaluates not only whether an agent avoids unsafe content, but also whether it remains useful, procedurally correct, and able to resolve legitimate user goals.

\item \textbf{Open evaluation roadmap.} HOB is designed as an open evaluation ecosystem rather than a closed benchmark. The most important next steps are to calibrate subjective juror scores against human expert judgment with an inter-rater reliability study, and to run an ablation that removes the curated traps, jurors, and audit rules so that the contribution of curated intelligence to objective failure detection can be isolated from the base Harness LLM. Beyond these, future work should expand the Red-Team Trap library, add domain-specific Juror Personas, strengthen audit rubrics for regulated settings, report confidence intervals over a denser configuration grid, and maintain released artifacts so that other teams can reproduce and extend the evaluation.
\end{itemize}

Overall, the findings position HOB as a scalable and cost-aware paradigm for AI agent evaluation. Its value is not only in the current experiments, but in the structure it provides for production evaluation. Human experts define reusable evaluation intelligence upfront. ProofAgent Harness executes that intelligence repeatedly. Smaller or local Harness LLMs can participate in asymmetric evaluation, reducing evaluation cost while preserving adversarial pressure, traceability, and evidence-linked reporting. This makes HOB a practical foundation for scalable, production-grade evaluation of agentic AI.

\section{Conclusion}
\label{sec:conclusion}

This paper introduced Human-on-the-Bridge (HOB), a scalable evaluation paradigm for agentic AI. HOB shifts human expertise from repeated manual review to upfront curation of reusable evaluation intelligence. This intelligence includes domain context, Red-Team Traps, Juror Personas, scoring guidelines, audit rules, and fallback policies. Once curated, ProofAgent Harness executes this intelligence repeatedly through multi-turn adversarial evaluations and produces evidence-linked reports.

The empirical evaluation showed that HOB can scale across domains, agent backbones, and Harness LLM tiers. Across 47 completed evaluation configurations, 470 run-level trials, and 23{,}500 agent turns, HOB surfaced behavioral and procedural failures that static benchmarks and single-evaluator scoring may miss. These included phantom tool-call claims, missing mandatory compliance actions, policy drift, manipulation paths, and safe but non-resolving refusals. The results also support a precise form of the asymmetric evaluation thesis: smaller or lower-cost Harness LLMs can still surface objective, trace-verifiable failures when guided by curated traps, juror perspectives, and audit rules, even though their subjective scores are more lenient than those of larger evaluators.

The central contribution is a shift in how AI agent evaluation is organized. HOB does not discard benchmarks, human review, LLM-as-judge methods, red teaming, trace auditing, or open evaluation infrastructure. It organizes them into a human-curated lifecycle. Human experts define evaluation intelligence upfront. ProofAgent Harness executes it at scale across repeated evaluations. This makes human judgment reusable, reduces dependence on continuous human intervention, and enables cost-aware evaluation for production-scale agent development.

HOB therefore reframes agent evaluation as an orchestration problem: how to combine human expertise, adversarial pressure, evaluator models, traces, rubrics, and reports into a repeatable evidence base. As agents become more autonomous and embedded in real workflows, evaluation must move beyond isolated scores toward scalable, auditable, and reusable evaluation systems. HOB provides one such foundation for production-grade, evidence-linked evaluation of agentic AI.

\section*{Acknowledgments}

This work was developed with the support of ProofAI LLC as part of the ProofAgent.ai open-source initiative. The author thanks the ProofAgent.ai community and early users for their feedback on AI agent evaluation, adversarial testing, and evidence-linked reporting.

\bibliographystyle{plain}
\bibliography{science_template}

\end{document}